\documentclass[aps,prl,twocolumn,showpacs,superscriptaddress,times,letterpaper]{revtex4}  
\usepackage{graphicx} 
\usepackage{dcolumn}  
\usepackage{bm}       
\usepackage{amssymb}  

\def\missET {{\not\!\! E_T}}

\begin{document}
\bibliographystyle{apsrev}

\title{Precision measurement of the top quark mass from dilepton events at CDF II}

\affiliation{Institute of Physics, Academia Sinica, Taipei, Taiwan 11529, Republic of China} 
\affiliation{Argonne National Laboratory, Argonne, Illinois 60439} 
\affiliation{Institut de Fisica d'Altes Energies, Universitat Autonoma de Barcelona, E-08193, Bellaterra (Barcelona), Spain} 
\affiliation{Baylor University, Waco, Texas  76798} 
\affiliation{Istituto Nazionale di Fisica Nucleare, University of Bologna, I-40127 Bologna, Italy} 
\affiliation{Brandeis University, Waltham, Massachusetts 02254} 
\affiliation{University of California, Davis, Davis, California  95616} 
\affiliation{University of California, Los Angeles, Los Angeles, California  90024} 
\affiliation{University of California, San Diego, La Jolla, California  92093} 
\affiliation{University of California, Santa Barbara, Santa Barbara, California 93106} 
\affiliation{Instituto de Fisica de Cantabria, CSIC-University of Cantabria, 39005 Santander, Spain} 
\affiliation{Carnegie Mellon University, Pittsburgh, PA  15213} 
\affiliation{Enrico Fermi Institute, University of Chicago, Chicago, Illinois 60637} 
\affiliation{Comenius University, 842 48 Bratislava, Slovakia; Institute of Experimental Physics, 040 01 Kosice, Slovakia} 
\affiliation{Joint Institute for Nuclear Research, RU-141980 Dubna, Russia} 
\affiliation{Duke University, Durham, North Carolina  27708} 
\affiliation{Fermi National Accelerator Laboratory, Batavia, Illinois 60510} 
\affiliation{University of Florida, Gainesville, Florida  32611} 
\affiliation{Laboratori Nazionali di Frascati, Istituto Nazionale di Fisica Nucleare, I-00044 Frascati, Italy} 
\affiliation{University of Geneva, CH-1211 Geneva 4, Switzerland} 
\affiliation{Glasgow University, Glasgow G12 8QQ, United Kingdom} 
\affiliation{Harvard University, Cambridge, Massachusetts 02138} 
\affiliation{Division of High Energy Physics, Department of Physics, University of Helsinki and Helsinki Institute of Physics, FIN-00014, Helsinki, Finland} 
\affiliation{University of Illinois, Urbana, Illinois 61801} 
\affiliation{The Johns Hopkins University, Baltimore, Maryland 21218} 
\affiliation{Institut f\"{u}r Experimentelle Kernphysik, Universit\"{a}t Karlsruhe, 76128 Karlsruhe, Germany} 
\affiliation{High Energy Accelerator Research Organization (KEK), Tsukuba, Ibaraki 305, Japan} 
\affiliation{Center for High Energy Physics: Kyungpook National University, Taegu 702-701, Korea; Seoul National University, Seoul 151-742, Korea; and SungKyunKwan University, Suwon 440-746, Korea} 
\affiliation{Ernest Orlando Lawrence Berkeley National Laboratory, Berkeley, California 94720} 
\affiliation{University of Liverpool, Liverpool L69 7ZE, United Kingdom} 
\affiliation{University College London, London WC1E 6BT, United Kingdom} 
\affiliation{Centro de Investigaciones Energeticas Medioambientales y Tecnologicas, E-28040 Madrid, Spain} 
\affiliation{Massachusetts Institute of Technology, Cambridge, Massachusetts  02139} 
\affiliation{Institute of Particle Physics: McGill University, Montr\'{e}al, Canada H3A~2T8; and University of Toronto, Toronto, Canada M5S~1A7} 
\affiliation{University of Michigan, Ann Arbor, Michigan 48109} 
\affiliation{Michigan State University, East Lansing, Michigan  48824} 
\affiliation{Institution for Theoretical and Experimental Physics, ITEP, Moscow 117259, Russia} 
\affiliation{University of New Mexico, Albuquerque, New Mexico 87131} 
\affiliation{Northwestern University, Evanston, Illinois  60208} 
\affiliation{The Ohio State University, Columbus, Ohio  43210} 
\affiliation{Okayama University, Okayama 700-8530, Japan} 
\affiliation{Osaka City University, Osaka 588, Japan} 
\affiliation{University of Oxford, Oxford OX1 3RH, United Kingdom} 
\affiliation{University of Padova, Istituto Nazionale di Fisica Nucleare, Sezione di Padova-Trento, I-35131 Padova, Italy} 
\affiliation{LPNHE, Universite Pierre et Marie Curie/IN2P3-CNRS, UMR7585, Paris, F-75252 France} 
\affiliation{University of Pennsylvania, Philadelphia, Pennsylvania 19104} 
\affiliation{Istituto Nazionale di Fisica Nucleare Pisa, Universities of Pisa, Siena and Scuola Normale Superiore, I-56127 Pisa, Italy} 
\affiliation{University of Pittsburgh, Pittsburgh, Pennsylvania 15260} 
\affiliation{Purdue University, West Lafayette, Indiana 47907} 
\affiliation{University of Rochester, Rochester, New York 14627} 
\affiliation{The Rockefeller University, New York, New York 10021} 
\affiliation{Istituto Nazionale di Fisica Nucleare, Sezione di Roma 1, University of Rome ``La Sapienza," I-00185 Roma, Italy} 
\affiliation{Rutgers University, Piscataway, New Jersey 08855} 
\affiliation{Texas A\&M University, College Station, Texas 77843} 
\affiliation{Istituto Nazionale di Fisica Nucleare, University of Trieste/\ Udine, Italy} 
\affiliation{University of Tsukuba, Tsukuba, Ibaraki 305, Japan} 
\affiliation{Tufts University, Medford, Massachusetts 02155} 
\affiliation{Waseda University, Tokyo 169, Japan} 
\affiliation{Wayne State University, Detroit, Michigan  48201} 
\affiliation{University of Wisconsin, Madison, Wisconsin 53706} 
\affiliation{Yale University, New Haven, Connecticut 06520} 
\author{A.~Abulencia}
\affiliation{University of Illinois, Urbana, Illinois 61801}
\author{J.~Adelman}
\affiliation{Enrico Fermi Institute, University of Chicago, Chicago, Illinois 60637}
\author{T.~Affolder}
\affiliation{University of California, Santa Barbara, Santa Barbara, California 93106}
\author{T.~Akimoto}
\affiliation{University of Tsukuba, Tsukuba, Ibaraki 305, Japan}
\author{M.G.~Albrow}
\affiliation{Fermi National Accelerator Laboratory, Batavia, Illinois 60510}
\author{D.~Ambrose}
\affiliation{Fermi National Accelerator Laboratory, Batavia, Illinois 60510}
\author{S.~Amerio}
\affiliation{University of Padova, Istituto Nazionale di Fisica Nucleare, Sezione di Padova-Trento, I-35131 Padova, Italy}
\author{D.~Amidei}
\affiliation{University of Michigan, Ann Arbor, Michigan 48109}
\author{A.~Anastassov}
\affiliation{Rutgers University, Piscataway, New Jersey 08855}
\author{K.~Anikeev}
\affiliation{Fermi National Accelerator Laboratory, Batavia, Illinois 60510}
% ...
\collaboration{CDF Collaboration\footnote{With visitors from $^a$University of Athens, 
$^b$University of Bristol, 
$^c$University Libre de Bruxelles, 
$^d$Cornell University, 
$^e$University of Cyprus, 
$^f$University of Dublin, 
$^g$University of Edinburgh, 
$^h$University of Heidelberg, 
$^i$Universidad Iberoamericana, 
$^j$University of Manchester, 
$^k$Nagasaki Institute of Applied Science, 
$^l$University de Oviedo, 
$^m$University of London, Queen Mary and Westfield College, 
$^n$Texas Tech University, 
$^o$IFIC(CSIC-Universitat de Valencia), 
}}
\noaffiliation

\begin{abstract}
We report a measurement of the top quark mass, $M_t$, in the dilepton decay channel of $t\bar{t}\rightarrow b\ell'^{+}\nu_{\ell'}\overline{b}\ell^{-}\overline{\nu}_{\ell}$ using an integrated luminosity of 1.0~fb$^{-1}$ of $p\bar{p}$ collisions collected with the CDF II detector. We apply a method that convolutes a leading-order matrix element with detector resolution functions to form event-by-event likelihoods; we have enhanced the leading-order description to describe the effects of initial-state radiation.  The joint likelihood is the product of the likelihoods from 78 candidate events in this sample, which yields a measurement of $M_{t} = 164.5 \pm 3.9(\textrm{stat.}) \pm 3.9(\textrm{syst.}) \mathrm{~GeV}/c^2$, the most precise measurement of $M_t$ in the dilepton channel.
\end{abstract}


\maketitle 

\date{\today}

The top quark, the weak isospin partner of the bottom quark, is the most massive of the known fundamental particles. The top quark mass, $M_t$, is a fundamental parameter in the standard model. Precise measurements of $M_t$ along with those of other standard model parameters can be used to place constraints on the mass of the Higgs boson~\cite{higgsconstr} and on particles in extensions to the standard model~\cite{topsusy}. Currently, the Tevatron collider at Fermilab is the only accelerator capable of producing top quarks, where they are primarily produced in pairs.  The dilepton channel, including decays with two charged leptons in the final-state ($t\bar{t}\rightarrow W^+bW^-\bar{b}\rightarrow b\ell'^{+}\nu_{\ell'}\overline{b}\ell^{-}\overline{\nu}_{\ell}$), has a small branching fraction but has the fewest jets in the final-state, giving a smaller dependence on the calibration of the jet energy scale and less ambiguity in jet-quark assignments.  Nevertheless, discrepancies between measurements in different decay channels could indicate contributions from physics beyond the standard model~\cite{newsources}.  
Previous measurements of $M_t$ in the dilepton channel~\cite{cdfdilep,d0dilep,Abulencia:2005uq}, while statistically limited, have yielded lower values than measurements in other decay channels~\cite{abulencia:022004, cdfrun1ljets,d0me,d0me2}.

The dilepton channel poses unique challenges in reconstructing the kinematics of $t\bar{t}$ events as two neutrinos from $W$ decays escape undetected. Measurements of $M_t$ in this channel made using Run I data~\cite{cdfdilep,d0dilep} and recent measurements made using Run II data~\cite{Abulencia:2006js} utilize methods that make a series of kinematic assumptions and integrate over the remaining unconstrained quantities.   The greatest statistical precision, however, was achieved through the application of a matrix-element method~\cite{kondo,dg,d0me} which makes minimal kinematic assumptions,  instead integrating the leading-order  matrix-element for $t\bar{t}$ production and decay over all unconstrained quantities.  The first application of this method to the dilepton channel by the CDF collaboration~\cite{Abulencia:2005uq,Abulencia:2006qj} used 340~pb$^{-1}$ of Run II data.

 This Letter reports a measurement using an enhanced version~\cite{jayatilaka} of the matrix-element method described in Ref.~\cite{Abulencia:2006qj}. The enhanced method accounts for initial-state radiation from the incoming partons and has substantially improved statistical power.  This measurement uses data collected by the CDF II detector between March 2002 and March 2006 corresponding to an integrated luminosity of 1.0~fb$^{-1}$ and includes the 340~pb$^{-1}$ used in Ref.~\cite{Abulencia:2006qj}.

The CDF II detector~\cite{Acosta:2004yw} is a general-purpose detector, designed to study $p\bar{p}$ collisions at the Tevatron collider. The charged particle tracking system consists of a silicon microstrip tracker and a drift chamber, both immersed in a 1.4~T magnetic field.  Electromagnetic and hadronic calorimeters surround the tracking system and measure particle energies. Drift chambers located outside the calorimeters detect muons. 

The data used in this measurement are collected using the same triggers as in Ref.~\cite{Abulencia:2006qj}. After events passing the trigger requirement are reconstructed, we impose the selection criteria defined as ``DIL'' in Ref.~\cite{Acosta:2004uw} to isolate the dilepton candidates.  These selection cuts yield 78 candidate events.

We express the probability density for $t\bar{t}$ decays as $P_s(\mathbf{x}|M_t)$, where $M_t$ is the top quark mass and $\mathbf{x}$ represents the lepton energy, jet energy, and $\missET$~\cite{metdef} measurements. We calculate $P_s(\mathbf{x}|M_t)$ using the theoretical description of the $t\bar{t}$ production and decay process with respect to $\mathbf{x}$; $P_s(\mathbf{x} | M_t)$ is proportional to the differential cross section, $d\sigma(M_t)/d\mathbf{x} $.

We evaluate $P_s(\mathbf{x}|M_t)$ by integrating over quantities that are not directly measured by the detector, such as neutrino momenta and quark energies.  While quark energies cannot be directly measured, they can be estimated from measured jet energies. We integrate over quark energies using a parameterized transfer function $W_{jet}(p,j)$, which is the probability of measuring jet energy $j$, given quark energy $p$.

 As in Ref.~\cite{Abulencia:2006qj}, we assume that lepton energies and quark angles are perfectly measured, that incoming partons are massless and have no transverse momentum, and that the two highest energy jets in the event correspond to the $b$ quarks from $t\bar{t}$ decay. Unlike in Ref.~\cite{Abulencia:2006qj}, we do not assume zero transverse momentum of the $t\bar{t}$ system, $p_T^{t\bar{t}}$, which would require no initial-state radiation. Instead, we infer likely values of  $p_T^{t\bar{t}}$ from unclustered transverse energy~\cite{uedef} and jets that are not the two most energetic in the event.   We parameterize the relation between these measured quantities and $p_T^{t\bar{t}}$ as a transfer function, $W_{p_T}(p_T^{t\bar{t}}, U)$, where $U$ is a sum of the unclustered transverse energy and sub-leading jet transverse energies in an event. Both $W_{jet}(p,j)$ and $W_{p_T}(p_T^{t\bar{t}}, U)$ are estimated using $t\bar{t}$ events generated with {\sc herwig}~\cite{herwig} and the CDF II detector simulation~\cite{cdfsim}.  This description of the initial state radiation improves the expected statistical uncertainty by 10\% compared to the technique described in Ref.~\cite{Abulencia:2006qj}.

The effect of the above assumptions on the final measurement is estimated using Monte Carlo simulation.  The expression for the probability density at a given mass for a specific event can be written as

\begin{widetext}
\begin{equation}
P_s(\mathbf{x}|M_t) = \frac{1}{N}\int d\Phi |{\mathcal M}_{t\overline{t}}(q_i,p_i;M_t)|^2 \prod_{k=1,2} W_{jet}(p_k,j_k) W_{p_T}(p_T^{t\bar{t}}, U) f_{PDF}(q_1)f_{PDF}(q_2), 
\label{eqn:sigprob}
\end{equation}
\end{widetext}
\noindent
where the integral $d\Phi$ is over the eight remaining unconstrained momenta of the initial and final-state particles, $q_1$ and $q_2$ are the incoming parton momenta, $p_i$ are the outgoing lepton and quark momenta, $f_{PDF}(q_i)$ are the parton distribution functions (PDFs)~\cite{cteq} and ${\mathcal M}_{t\overline{t}}(q_i,p_i;M_t)$ is the leading-order $t\overline{t}$ production and decay matrix element as defined in Refs.~\cite{meref1, meref2} for the process $q\overline{q} \rightarrow t\overline{t} \rightarrow b\ell^{+}\nu_\ell\overline{b}\ell'^{-}\overline{\nu}_{\ell'}$~\cite{glueglue}. The term $1/N$ is defined such that the probability density satisfies the normalization condition, $\int d\mathbf{x}\ P_s(\mathbf{x}|M_t) = 1$. The probability for both possible jet-parton assignments is evaluated and summed. 

In addition to $t\bar{t}$ production, we calculate the probability for dominant background processes. The final event-by-event probability is then $P(\mathbf{x}|M_t) = P_{s}(\mathbf{x}|M_t) p_s + P_{b_1}(\mathbf{x}) p_{b_1} + P_{b_2}(\mathbf{x}) p_{b_2}\cdots$, where $p_s$ and $p_{b_i}$ are determined from the expected fractions of signal and background events (see Table~\ref{tab:bkg}). To determine the $P_{b_i}$, we numerically evaluate background matrix elements using algorithms adopted from the {\sc alpgen}~\cite{alpgen} generator. We calculate probabilities for the following background processes: $Z/\gamma^* \to ee,\mu\mu$ plus associated jets, $W+\ge 3$ jets where one jet is incorrectly identified as a lepton, and $WW$ plus associated jets. We do not calculate probabilities for $Z\to\tau\tau$ or $WZ$, comprising 11\% of the expected background. Studies indicate that use of the background probabilities improves the expected statistical uncertainty by 10\%. The posterior probability for the sample is the product of the event-by-event probabilities. The mean of the posterior probability, $P(M_t)$, is the raw measured mass, $M_t^{raw}$, and its standard deviation is the raw measured statistical uncertainty, $\Delta M_t^{raw}$.  Both are subject to corrections, described below.

\begin{table}[h]
\caption {Expected numbers of signal and background events for a data sample of integrated luminosity of 1.0~fb$^{-1}$. The number of expected $t\bar{t}$ is given for $\sigma_{t\bar{t}}=9.1$~pb, which corresponds to $M_t=165$~GeV$/c^2$. Other backgrounds are negligible; expected signal and background numbers have an additional correlated uncertainty of 6\% from uncertainty in the sample luminosity. }
\begin{ruledtabular}
\begin{tabular}{lr}
Source & Events \\
\hline
Expected $t\overline{t}$~($M_t=165$~GeV/$c^2$) &$63.4\pm 1.7$ \\
 \hline 
 Expected Background & $26.9\pm 4.8$ \\
\ \ \  Drell-Yan ($Z/\gamma^*\to\ell\ell$) & $13.1\pm 4.4$\ \ \ \ \ \ \ \ \ \ \\
\ \ \  Misidentified Lepton & $8.7\pm 1.5$\ \ \ \ \ \ \ \ \ \ \\
\ \ \  Diboson ($WW/WZ$) & $5.1\pm 1.0$\ \ \ \ \ \ \ \ \ \ \\[6pt]
\hline 
 Total Expected ($M_t=165$~GeV/$c^2$) & $90.3\pm 5.1$ \\
 \hline 
 Run II Observed & 78 \\
\end{tabular}

\end{ruledtabular}
\label{tab:bkg}\rm
\end{table}

\begin{figure}[ht]
\includegraphics[width=1.0\columnwidth]{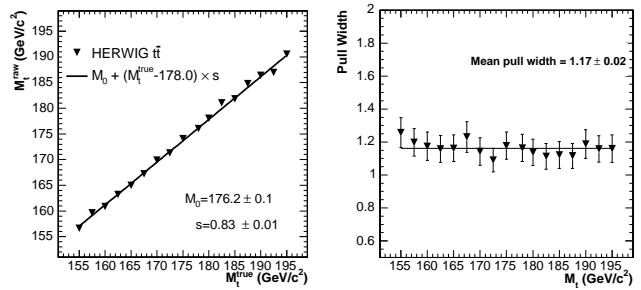}
\caption{Left: Mean measured $M_t$ in Monte Carlo experiments of signal and background events at varying top quark mass. The solid line is a linear fit to the points. Right: Pull widths of Monte Carlo experiments of signal and background events at varying top quark mass. The solid line is the average of all points, $1.17 \pm 0.02$.}
\label{fig:response}
\end{figure}

To test the performance of our method we perform Monte Carlo experiments of signal and background events. Signal events are generated using {\sc herwig} for top quark masses ranging from 155~GeV$/c^2$ to 195~GeV$/c^2$. Background events are modeled using observed events in the case of background due to misidentified leptons, {\sc alpgen}-simulated events in the case of $Z/\gamma^*\to ee,\mu\mu$, and {\sc pythia}-simulated~\cite{pythia} events in the case of $Z/\gamma^*\to \tau\tau, WW,WZ$, and $ZZ$. The numbers of signal and background events in each Monte Carlo experiment are Poisson-fluctuated values around the mean values given in Table~\ref{tab:bkg}. The estimate for the $t\bar{t}$ signal at varying masses is evolved to account for the variation of cross-section and acceptance. The response of the method for these Monte Carlo experiments is shown in Fig.~\ref{fig:response} (left). While the response is consistent with a linear dependence on the top quark mass, its slope is less than unity due to the presence in the sample of background events for which probabilities are not calculated. Corrections, $ M_t = 178 \textrm{~GeV}/c^2 + (M_t^{raw}-176.4 \textrm{~GeV}/c^2)/0.83$ and $\Delta M_t = \Delta M_t^{raw}/0.83 $, are derived from this response and applied to values measured in the data.

The width of the pull distributions in these Monte Carlo experiments, shown in Fig.~\ref{fig:response} (right), where pull is defined as $(M_t-M_t^{true})/\Delta M_t$, indicates that the statistical uncertainty is underestimated by a factor of 1.17, after applying the corrections described above. This results from the simplifying assumptions described above, made to ensure the computational tractability of the integrals in Eq.~\ref{eqn:sigprob}.  The largest effects~\cite{Abulencia:2006qj} are the leading two jets in an event not resulting from $b$-quark hadronization, imperfect lepton momentum resolution, imperfect jet angle resolution, and unmodeled backgrounds.  Correcting by this factor of 1.17, we estimate the mean statistical uncertainty to be 5.0$\mathrm{~GeV}/c^2$ if $M_t=175 \mathrm{~GeV}/c^2$ or $4.2 \mathrm{~GeV}/c^2$ if $M_t=165 \mathrm{~GeV}/c^2$.

\begin{figure}[ht]
\includegraphics[width=1.0\columnwidth]{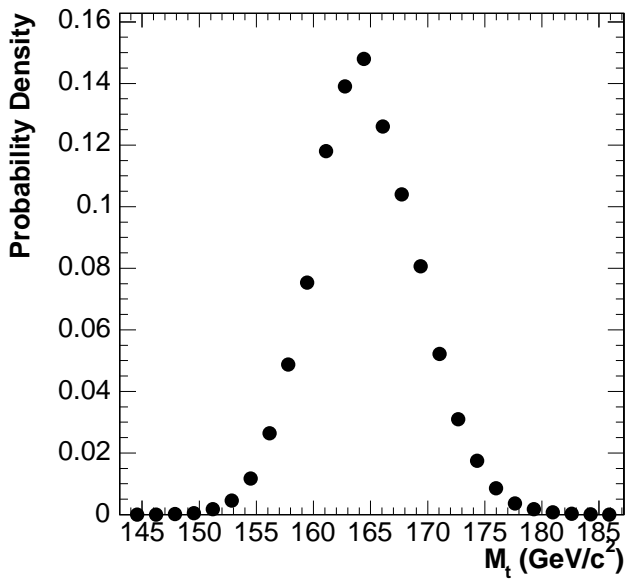}
\caption{Joint posterior probability density as a function of the top quark mass for the 78 observed candidate events, after all corrections. Systematic uncertainties are not shown.} 
\label{fig:dataprob}
\end{figure}

Applying the method and corrections described above to the 78 candidate events observed in the data, we measure $M_{t}=164.5\pm 3.9 (\textrm{stat.}) \textrm{ GeV}/c^2$. Figure~\ref{fig:dataprob} shows the joint probability density, without systematic uncertainty, for the events in our data set.

The measured statistical uncertainty is consistent with the distribution of statistical uncertainties in Monte Carlo experiments where signal events with $M_t = 165\textrm{ GeV}/c^2$ are chosen according to a Poisson distribution with mean $N_{t\overline{t}} = 63.4$ events. This number of events corresponds to the cross section and acceptance at $M_t = 165\textrm{ GeV}/c^2$. Of these Monte Carlo experiments, 31\% yielded a statistical uncertainty less than 3.9 GeV$/c^2$.

\begin{table}
\caption{Summary of systematic uncertainties.}
\begin{ruledtabular}
\begin{tabular}{lc}

Source & $\Delta M_t$ (GeV$/c^2$)\\\hline
Jet energy scale & 3.5 \\
Limited background statistics & 0.7 \\
PDFs & 0.8 \\
Generator & 0.9 \\
Background modeling & 0.2 \\
FSR modeling & 0.3 \\
ISR modeling & 0.3 \\
Response correction & 0.6 \\
Sample composition uncertainty & 0.7 \\
Lepton energy scale & 0.1 \\[6pt]
Total &  3.9\\
\end{tabular}

\end{ruledtabular}
\label{tab:sys}\rm
\end{table}

A summary of systematic uncertainties in this measurement is shown in Table~\ref{tab:sys}. The largest source of systematic uncertainty in our measurement is due to uncertainty in the jet energy scale~\cite{jes}, which we estimate at 3.5~GeV$/c^2$ by varying the scale within its uncertainty, including effects of high instantaneous luminosity (which have been found to contribute an uncertainty of 0.2~GeV$/c^2$).  This is necessarily larger than in the previous application of this method~\cite{Abulencia:2006qj}, as we have included additional jets measurements in our calculation; future measurements would benefit from a direct calibration of the $b$-jet energy scale from $Z\rightarrow b\overline{b}$ decays. We estimate the uncertainty due to the limited number of background events available for Monte Carlo experiments to be 0.7~GeV$/c^2$. Uncertainties due to PDFs are estimated using different PDF sets (CTEQ5L~\cite{cteq} vs. MRST72~\cite{mrst}), different values of $\Lambda_{QCD}$ and varying the eigenvectors of the CTEQ6M~\cite{cteq} set; the quadrature sum of these uncertainties is 0.8~GeV$/c^2$. Uncertainty due to showering model in the Monte Carlo generator used for $t\bar{t}$ events is estimated as the difference in the extracted top quark mass from {\sc pythia} events and {\sc herwig} events and amounts to 0.9~GeV$/c^2$. We estimate the uncertainty coming from  modeling of the two largest sources of background, $Z/\gamma^*$ and events with a misidentified lepton, to be 0.2~GeV$/c^2$. Uncertainty due to imperfect modeling of initial-state (ISR) and final state (FSR) QCD radiation is estimated by varying the amounts of ISR and FSR in simulated events~\cite{ljetsprd}, giving 0.3~GeV$/c^2$ for FSR and 0.3~GeV$/c^2$ for ISR. The uncertainty in the mass due to uncertainties in the response correction shown in Fig.~\ref{fig:response} is 0.6~GeV$/c^2$.  The contribution from uncertainties in background composition is estimated by varying the background estimates from Table~\ref{tab:bkg} within their uncertainties and amounts to 0.7 GeV$/c^2$. The uncertainty in the lepton energy scale contributes an uncertainty of 0.1~GeV$/c^2$ to our measurement. Adding all of these contributions together in quadrature yields a total systematic uncertainty of 3.9 GeV/$c^2$.

 In summary, we have presented a new measurement of the top quark mass in the dilepton channel, $M_t=164.5\pm 3.9 (\textrm{stat.}) \pm 3.9 (\textrm{syst.})$~GeV$/c^2$. This is the most precise measurement of $M_t$ in this channel with an approximately 35\% improvement in statistical precision over the previous best measurement~\cite{Abulencia:2006qj}. The systematic uncertainly,  while 15\% larger, is nearly completely correlated with systematic uncertainties in measurements in other channels and so does not impact the global combination nor an analysis of measurements in different channels.  Previous measurements yielded smaller values of $M_t$ in the dilepton channel~\cite{cdfdilep,d0dilep,Abulencia:2005uq} than in the single lepton~\cite{abulencia:022004} and all-hadronic~\cite{allhadprl} decay channels, though the discrepancy was not statistically significant. Our measurement continues that trend with substantially increased statistical precision.   A global combination~\cite{worldave}, however, shows that these variations are consistent with statistical fluctuations. 

We thank the Fermilab staff and the technical staffs of the participating institutions for their vital contributions. This work was supported by the U.S. Department of Energy and National Science Foundation; the Italian Istituto Nazionale di Fisica Nucleare; the Ministry of Education, Culture, Sports, Science and Technology of Japan; the Natural Sciences and Engineering Research Council of Canada; the National Science Council of the Republic of China; the Swiss National Science Foundation; the A.P. Sloan Foundation; the Bundesministerium f\"ur Bildung und Forschung, Germany; the Korean Science and Engineering Foundation and the Korean Research Foundation; the Particle Physics and Astronomy Research Council and the Royal Society, UK; the Institut National de Physique Nucleaire et Physique des Particules/CNRS; the Russian Foundation for Basic Research; the Comisi\'on Interministerial de Ciencia y Tecnolog\'{\i}a, Spain; the European Community's Human Potential Programme under contract HPRN-CT-2002-00292; and the Academy of Finland.

\bibliography{prl}

\begin{thebibliography}{32}
\bibitem[{met()}]{metdef}
\bibinfo{note}{Missing transverse energy, $\missET$, is defined as the
  magnitude of the vector, $ -\sum_i E_T^i \vec n_i$, where $E_T^i$ are the
  magnitudes of transverse energy contained in each calorimeter tower $i$, and
  $\vec n_i$ is the unit vector from the interaction vertex to the tower in the
  transverse ($x,y$) plane.}
\bibitem[{ued()}]{uedef}
\bibinfo{note}{The unclustered transverse energy in an event is the total
  transverse energy in the event that is measured in the calorimeter but not
  clustered into a lepton or jet.}
\bibitem[{glu()}]{glueglue}
\bibinfo{note}{While up to 15\% of $t\overline{t}$ pairs at the Tevatron are
  produced by gluon-gluon fusion ($gg\to t\overline{t}$), this term can be
  excluded from the matrix element with negligible effect on the precision of
  the measurement.}
\end{thebibliography}

\end{document}